\documentclass[twocolumn,showpacs,prseparableeprintnumbers,amsmath,amssymb]{revtex4}
\usepackage{graphicx}% Include figure files
\usepackage{dcolumn}% Align table columns on decimal point
\usepackage{bm}% bold math

\begin{document}

%\preprint{APS/123-QED}

\title{Optimal estimation of an observable's expectation value for pure states\\
for general measure of deviation}% Force line breaks with \\

\author{Minoru Horibe}
  \email{horibe@quantum.apphy.fukui-u.ac.jp}
\author{Akihisa Hayashi}
 \email{hayashi@soliton.apphy.fukui-u.ac.jp}
\author{Takaaki Hashimoto}
 \email{d901005@icpc00.icpc.fukui-u.ac.jp}
\affiliation{
Department of Applied Physics, University of Fukui, Fukui 910-8507,Japan}%

\date{\today}% It is always \today, today,

\begin{abstract}
 We investigate the optimal estimation of quantum expectation value of a physical observable, which minimizes a mean error with respect to general measure of 
deviation, when a finite number of copies of a pure state are prepared. 
If pure sates are uniformly distributed, the minimum value of mean error for any measure of deviation
is achieved by projective measurement on each copy.
\end{abstract}

\maketitle

%%%%%%%%%%%%%%%%%%%% Introduction %%%%%%%%%%%%%%%%%%%%%%%%%%%%%%%%%
\section{\label{sec:level1}Introduction}
In the quantum information theory, it is one of important issue to pick out the best way
how to extract information from given quantum state.
Many problems related to this have been investigated for a long time since Helstrom's and Holevo's works \cite{Helst,Holevo}. In these problems, many copies of quantum state are needed, because
we can get few information from one state, owing to statistical properties of quantum mechanics.
The measurements on these copies are classified into two types. One is separable measurement whose
positive operator valued measure(POVM) is described by tensor products of POVM on each copy, the other is joint measurements whose
POVM includes elements with entangled eigenvectors. 

In many cases
like quantum state estimation \cite{Massar95,Derka98,Bruss99,chirib042,hayashi1},
 sending some information by qubits \cite{Bag101,Bag201,chirib041} and state identification \cite{hayashi2,hayashi3}, 
it
seems that the joint measurement is optimal.
For example, in the paper \cite{hayashi2}, Hayashi et al. discussed a problem of identifying a given pure state with one of two reference pure states, when no classical knowledge on the reference states is given, but a certain number of copies of them are presented. It was shown that the averaged success
probability takes the maximum when we adopt joint measurement on tensor product of reference states and unknown state. However, Masahito Hayashi \cite{Mhayashi} investigated an estimation with minimum mean error with respect to general measure of deviation which satisfies appropriate conditions and showed that the minimum mean error over all
measurement coincides with minimum mean error over separable measurement when the number of copies goes to 
infinity.
For different situations \cite{Bagan05,Ball04,Giov06}, the similar result was obtained.

In the paper \cite{DAGiPer}, D'Ariano et al. discussed the optimal estimation 
of an observable's expectation value for the finite copies of pure-state and showed that the measurement which minimizes the variance is separable one if POVM is unbiased namely averaged estimator over the repeated measurement is equal to the expectation value.
In the previous paper \cite{hayashi4}, when $N$ copies of pure state distributed uniformly in $d$-dimensional Hilbert space is presented, it was seen that the measurement which minimizes the averaged variance over all pure state is separable measurement on each copy of $N$ pure states.
Thus for the estimation of observable's expectation value, it seems that separable measurement is optimal. In this paper, in order to make sure of this statement, we investigate whether  
minimum value of mean error with respect to any measure of deviation
is achieved by separable measurement of each state or not.  

%%%%%%%%%%%%%%%%%%%%% Estimation for ... %%%%%%%%%%%%%%%%%%%%%%%%%%
\section{\label{sec:opt} Estimation with minimum mean error for general measure of deviation}
In the previous paper \cite{hayashi4}, we determined the optimal way to estimate expectation value
of a physical observable $\Omega$, namely,  POVM and estimator for each element of it which 
minimize the variance, when $N$ copies of unknown pure state $|\phi\rangle$
on $d$-dimensional Hilbert space are prepared. In the same situation  we find the POVM $\{E_a\}$
and estimator $\omega_a$ which minimize the mean error $\langle \Delta \rangle$ defined by,
\begin{equation}
\langle \Delta \rangle=\left\langle \sum_{a=1}
{\rm Tr}[\rho^{\otimes N} E_a]W(\omega_a-{\rm Tr}[\Omega\rho]) \right\rangle,
\label{defdev}
\end{equation}
where $\rho$ is density matrix for pure sate $|\phi\rangle$,
\[
\rho=|\phi\rangle\langle\phi|,
\]
bracket $\langle \cdots \rangle$ means the average for $|\phi \rangle$ which is distributed uniformly over the $d$-dimensional Hilbert space and the function $W(x)$ is a general measure of deviation. For simplicity, we assume that this function satisfies the conditions;
\begin{itemize}
\item[(a)] $W(x)>0\;\;(x\neq 0)$ and $W(0)=0$,
\item[(b)] $\frac{dW(x)}{dx}>0\;\;(x>0)$, $\frac{dW(x)}{dx}<0\;\;(x<0)$
\item[(c)] $\frac{d^2W(x)}{dx^2}>0$
\end{itemize} 
We will discuss the case where  $W(x)$ satisfies weaker conditions in the next section.

Expanding the pure state $|\phi \rangle$,
\[
|\phi \rangle=\sum_{n=1}^d c_n|n\rangle ,
\]
in the eigenvectors $|n\rangle,\;\;(n=1,2,\cdots d)$ with eigenvalue $\lambda_n$ of the observable $\Omega$,
\[
\Omega |n \rangle = \lambda_n|n \rangle \;\;(n=1,2,\cdots, d),
\]
the density matrix $\rho^{\otimes N}$ becomes
\begin{eqnarray}
\rho^{\otimes N}&=&
        \sum_{n_1=1}^d\cdots\sum_{n_N=1}^d 
        \sum_{m_1=1}^d\cdots\sum_{m_N=1}^d
        c_{n_i}c^{\ast}_{m_i}\cdots c_{n_N}c^{\ast}_{m_N} \nonumber\\
&&\times |n_1\rangle\langle m_1|\otimes \cdots \otimes |n_N\rangle\langle m_N|, \nonumber\\
&=&\sum_{{\tiny \begin{array}{c}
                          s_1,\dots,s_d\ge 0\\
                          s_1+\cdots+s_d=N
                           \end{array}}}
   \sum_{{\tiny \begin{array}{c}
                          s'_1,\dots,s'_d\ge 0\\
                          s'_1+\cdots+s'_d=N
                           \end{array}}}\nonumber\\
&&\times c_1^{s_1}\cdots c_d^{s_d}(c^{\ast}_1)^{s'_1}\cdots(c^{\ast}_d)^{s'_d}\nonumber\\
&&\times \sum_{{\tiny \begin{array}{c}
                          \{n_1,\cdots,n_N\}\\
                          (s_1,\cdots,s_d)
                          \end{array}}}
   \sum_{{\tiny \begin{array}{c}
                          \{m_1,\cdots,m_N\}\\
                         (s'_1,\cdots,s'_d)
                          \end{array}}}\nonumber\\
&&\times |n_1\rangle\langle m_1|\otimes \cdots \otimes |n_N\rangle\langle m_N| .
\label{genfdev}
\end{eqnarray}
Here, $s_n\;(n=1,2,\cdots,d)$ is occupation number of the eigenstate
$|n\rangle \;(n=1,2,\cdots,d)$ so 
the notation $\displaystyle{\sum_{{\tiny \begin{array}{c}
                          s_1,\dots,s_d\ge 0\\
                          s_1+\cdots+s_d=N
                           \end{array}}}}$
 means that the summation is taken over all non-negative integers
$s_l\;\;(l=1,2,\cdots,d)$ which satisfy the condition $s_1+s_2+\cdots + s_d=N$ and
the
notation $\displaystyle{\sum_{{\tiny \begin{array}{c}
                          \{n_1,\cdots,n_N\}\\
                          (s_1,\cdots,s_d)
                          \end{array}}}}$
 means that the summation is taken over all states with same occupation number $s_l\;(l=1,2,\cdots,d)$.

Substituting this expression for the density matrix $\rho^{\otimes N}$ into the definition(\ref{defdev}) of $\langle \Delta \rangle$  we want to minimize, we have 
\begin{eqnarray}
\langle \Delta \rangle&=&\frac{1}{Z}\sum_{a}\int\prod_{n=1}^ddc_ndc_n^{\ast}
\delta\left(\sum_{n=1}^d|c_n|^2-1\right)\nonumber\\
&&\times \sum_{{\tiny \begin{array}{c}
                          s_1,\dots,s_d\ge 0\\
                          s_1+\cdots+s_d=N
                           \end{array}}}
   \sum_{{\tiny \begin{array}{c}
                          s'_1,\dots,s'_d\ge 0\\
                          s'_1+\cdots+s'_d=N
                           \end{array}}}\nonumber\\
&&\times c_1^{s_1}\cdots c_d^{s_d}(c^{\ast}_1)^{s'_1}\cdots(c^{\ast}_d)^{s'_d}\nonumber\\
&&\times \sum_{{\tiny \begin{array}{c}
                          \{n_1,\cdots,n_N\}\\
                          (s_1,\cdots,s_d)
                          \end{array}}}
   \sum_{{\tiny \begin{array}{c}
                          \{m_1,\cdots,m_N\}\\
                         (s'_1,\cdots,s'_d)
                          \end{array}}}\nonumber\\
&&\times |n_1\rangle\langle m_1|\otimes \cdots \otimes |n_N\rangle\langle m_N|\nonumber\\
&&\times W\left(\omega_a-\sum_{n=1}^d\lambda_{n}|c_n|^2\right),
\end{eqnarray}
where $dc_ndc_n^{\ast}$ stands for the integration over real part $c_{n{\rm R}}$ and imaginary part
$c_{n{\rm I}}$ of complex number $c_n$ and $Z$ is normalisation factor,
\begin{eqnarray*}
Z&=&\int\prod_{n=1}^ddc_ndc_n^{\ast}
\delta\left(\sum_{n=1}^d|c_n|^2-1\right)\\
&=&\int\prod_{n=1}^ddc_{n{\rm R}}dc_{n{\rm I}}
\delta\left(\sum_{n=1}^d|c_n|^2-1\right)=\frac{1}{2}\frac{2\pi^{d}}{\Gamma\left(d\right)}.
\label{exdefdev}
\end{eqnarray*}

Changing variables of integration $c_n$ to $\xi_n$ and $\varphi_n$,
\[
 c_n=\xi_ne^{i\varphi_n}\;\;(n=1,2,\cdots, d),
\]
the mean error $\langle \Delta \rangle$ becomes
\begin{eqnarray*}
\langle \Delta \rangle&=&\frac{1}{Z}\sum_{a}
\sum_{{\tiny \begin{array}{c}
                          s_1,\dots,s_d\ge 0\\
                          s_1+\cdots+s_d=N
                           \end{array}}}
\sum_{{\tiny \begin{array}{c}
                          s'_1,\dots,s'_d\ge 0\\
                          s'_1+\cdots+s'_d=N
                           \end{array}}} \\
&\times&\biggr[
\int^{2\pi}_{0}\left(\prod_{n=1}^dd\varphi_n\right)
e^{i\{(s_1-s'_1)\varphi_1+\cdots+(s_d-s'_d)\varphi_d\}} \\
&\times&\sum_{{\tiny \begin{array}{c}
                          \{n_1,\cdots,n_N\}\\
                          (s_1,\cdots,s_d)
                          \end{array}}}
\sum_{{\tiny \begin{array}{c}
                          \{m_1,\cdots,m_N\}\\
                          (s'_1,\cdots,s'_d)
                          \end{array}}}\int^{\infty}_{0}
\left(\prod_{n=1}^{d}d\xi_n\right)\nonumber\\
&\times&\xi_1^{s_1+s'_1+1}\cdots\xi_d^{s_d+s'_d+1}
\delta\left(\sum_{n=1}^d\xi^2_n-1\right) \\
&\times& 
{\rm Tr}[|n_1\rangle\langle m_1| \otimes \cdots  \otimes |n_N\rangle\langle m_N| E_a]\nonumber\\
&\times&W\left(\omega_a-\sum_{n=1}^d\lambda_{n}\xi_n^2 \right)\biggr].
\end{eqnarray*}
The contribution of the term where $s_n$ is different from $s'_n$ vanishes under the $d\varphi_n\;\;(n=1,2,\cdots,d)$ integration,
and we have
\begin{equation}
\langle \Delta \rangle=\sum_{{\tiny \begin{array}{c}
                          s_1,\dots,s_d\ge 0\\
                          s_1+\cdots+s_d=N
                           \end{array}}}
w_{s_1,\cdots,s_d}(\omega_a)
{\rm Tr}[{\cal{M}}_{{s_1} \cdots {s_n}} E_a],
\label{exdefdev2}
\end{equation}
where the function $w_{s_1,\cdots,s_d}(\omega_a)$ and operator ${\cal{M}}_{{s_1} \cdots {s_n}}$ are defied by
\begin{eqnarray}
w_{s_1,\cdots,s_d}(\omega_a)&=&\frac{(2\pi)^d}{Z}\int_0^{\infty}
\prod_{n=1}^dd\xi_n
\delta\left(\sum_{n=1}^d\xi_n^2-1\right) \nonumber\\
&\times& \prod_{n=1}^d\xi_n^{2s_n+1} W\left(\omega_a-\sum_{n=1}^d\lambda_{n}\xi_n^2\right) \label{deffuw}\\
{\cal{M}}_{{s_1} \cdots {s_n}}&=&\sum_{{\tiny \begin{array}{c}
                          \{n_1,\cdots,n_N\}\\
                          (s_1,\cdots,s_d)
                          \end{array}}}
                          {\sum_{{\tiny \begin{array}{c}
                          \{m_1,\cdots,m_N\}\\
                          (s_1,\cdots,s_d)
                          \end{array}}}},\nonumber\\
&\times&
|n_1\rangle\langle m_1| \otimes \cdots  \otimes |n_N\rangle\langle m_N|.
\label{defmam}
\end{eqnarray}
Clearly, the operator ${\cal{M}}_{{s_1} \cdots {s_n}}$  is non-negative and its rank is equal to one; 
\begin{eqnarray*}
{\cal{M}}_{{s_1} \cdots {s_n}}
&=&\left(\sum_{{\tiny \begin{array}{c}
                          \{n_1,\cdots,n_N\}\\
                          (s_1,\cdots,s_d)
                          \end{array}}}
  |n_1\rangle \otimes \cdots \otimes |n_N \rangle\right)\\
  &\times& \left(\sum_{{\tiny \begin{array}{c}
                          \{m_1,\cdots,m_N\}\\
                          (s_1,\cdots,s_d)
                          \end{array}}}
  \langle m_1|\otimes \cdots \otimes \langle m_N| \right).
\end{eqnarray*}

From the conditions (a) $\sim$ (c) for $W(x)$,
we can see that the first derivative of function $w_{s_1,\cdots,s_d}(\omega_a)$ becomes 
positive(negative)  as $x$ goes to $+\infty(-\infty)$
and that the second derivative of it is positive. Thus there is only one value
$\Omega^{({\rm min})}_{s_1,\cdots,s_d}$ where the first derivative of the function
$w_{s_1,\cdots,s_d}(\omega_a)$ vanishes and the function $w_{s_1,\cdots,s_d}(\omega_a)$ 
takes the minimum.  
As the operator ${\cal{M}}_{{s_1} \cdots {s_n}}$ is non-negative, we can get lower limit of
mean error $\langle \Delta \rangle$,
\begin{eqnarray}
\langle \Delta \rangle
&=&
\sum_{{\tiny \begin{array}{c}
                          s_1,\dots,s_d\ge 0\\
                          s_1+\cdots+s_d=N
                           \end{array}}}
w_{s_1,\cdots,s_d}(\omega_a)
{\rm Tr}[{\cal{M}}_{{s_1} \cdots {s_d}} E_a], \nonumber\\
&\ge&
\sum_{{\tiny \begin{array}{c}
                          s_1,\dots,s_d\ge 0\\
                          s_1+\cdots+s_d=N
                           \end{array}}}
w_{s_1,\cdots,s_d}(\Omega^{({\rm min})}_{s_1,\cdots,s_d}) \nonumber\\
&\times& {\rm Tr}[{\cal{M}}_{{s_1} \cdots {s_n}} E_a], \nonumber\\
&=&\sum_{{\tiny \begin{array}{c}
                          s_1,\dots,s_d\ge 0\\
                          s_1+\cdots+s_d=N
                           \end{array}}}
w_{s_1,\cdots,s_d}(\Omega^{({\rm min})}_{s_1,\cdots,s_d}) \nonumber\\
&\times&{\rm Tr}\left[{\cal{M}}_{{s_1} \cdots {s_n}} \left(\sum_{a}E_a\right)\right], \nonumber\\
&=&\sum_{{\tiny \begin{array}{c}
                          s_1,\dots,s_d\ge 0\\
                          s_1+\cdots+s_d=N
                           \end{array}}}
w_{s_1,\cdots,s_d}(\Omega^{({\rm min})}_{s_1,\cdots,s_d})\nonumber\\
&\times&{\rm Tr}\left[{\cal{M}}_{{s_1} \cdots {s_n}} \right], 
\label{fnlineq}
\end{eqnarray}
Now,  
we prove that this lower limit is achieved if we choose ${\cal P}_{s_1\cdots,s_d}$ which is defined by
\begin{equation}
{\cal{P}}_{s_1\cdots,s_d}=
\sum_{{\tiny \begin{array}{c}
                          \{n_1,\cdots,n_N\}\\
                          (s_1,\cdots,s_d)
                          \end{array}}}
|n_1\rangle\langle n_1|\otimes \cdots \otimes |n_N\rangle\langle n_N|,
\label{defPOVM}
\end{equation}
and which satisfies the condition for projective operators
\begin{eqnarray*}
&&\sum_{{\tiny \begin{array}{c}
                          \{n_1,\cdots,n_N\}\\
                          (s_1,\cdots,s_d)
                          \end{array}}}{\cal{P}}_{s_1\cdots,s_d}={\bf 1},\\
&&{\cal{P}}_{s_1\cdots,s_d}{\cal{P}}_{s'_1\cdots,s'_d}=\delta_{s_1s'_1}\cdots
\delta_{s_ds'_d}{\cal{P}}_{s_1\cdots,s_d}.
\end{eqnarray*}
as the POVM 
and $\Omega^{({\rm min})}_{s_1,\cdots,s_d}$ as the estimator for each element of this POVM.

 Under this choice of POVM and 
 estimators, the mean error $\langle \Delta \rangle$ in equation (\ref{exdefdev2}) becomes   
\begin{eqnarray*}
\langle \Delta \rangle&=&
\sum_{{\tiny \begin{array}{c}
                          s'_1,\dots,s'_d\ge 0\\
                          s'_1+\cdots+s'_d=N
                           \end{array}}}
\sum_{{\tiny \begin{array}{c}
                          s_1,\dots,s_d\ge 0\\
                          s_1+\cdots+s_d=N
                           \end{array}}}\\
&\times& w_{s_1,\cdots,s_d}(\Omega^{({\rm min})}_{s'_1,\cdots,s'_d})
{\rm Tr}[{\cal{M}}_{{s_1} \cdots {s_n}} {\cal{P}}_{s'_1,\cdots,s'_d}]\\
&=&
\sum_{{\tiny \begin{array}{c}
                          s_1,\dots,s_d\ge 0\\
                          s_1+\cdots+s_d=N
                           \end{array}}}\\
&\times& w_{s_1,\cdots,s_d}(\Omega^{({\rm min})}_{s_1,\cdots,s_d})
{\rm Tr}[{\cal{M}}_{{s_1} \cdots {s_n}}],
\end{eqnarray*}
since ${\cal P}_{s_1\cdots,s_d}$ and ${\cal{M}}_{{s_1} \cdots {s_n}}$ satisfy the following relation
\begin{eqnarray*}
{\cal{M}}_{{s_1} \cdots {s_n}}{\cal{P}}_{s'_1\cdots,s'_d}&=&
{\cal{P}}_{s'_1\cdots,s'_d}{\cal{M}}_{{s_1} \cdots {s_d}}\\
&=&\delta_{s_1s'_1}\cdots
\delta_{s_ds'_d}
{\cal{M}}_{{s_1} \cdots {s_n}}. 
\end{eqnarray*}

Thus we can obtain the POVM $\{{\cal P}_{s_1,\cdots,s_d}\}$ and estimators
$\Omega^{({\rm min})}_{s_1,\cdots,s_d}$ which minimizes the mean error $\langle \Delta \rangle$ for any measure of deviation
. Eigenvectors of some elements of this POVM are entangled  
states of $N$ copies (i.e any linear combination of states with same occupation number) and the measurement for this POVM is joint one.  However, in the light of the facts that
elements of this POVM are linear combination of the projection operators
$|n_1\rangle \langle n_1| \otimes \cdots \otimes |n_N \rangle \langle n_N|$ 
and that the mean error is linearly dependent on the element of POVM, we have another optimal measurement
described by projection operators $P_{n_1\cdots n_N}\;(n_1,\cdots,n_N=1,2,\cdots,d)$
\[
P_{n_1\cdots n_N}=|n_1\rangle \langle n_1| \otimes \cdots \otimes |n_N \rangle \langle n_N|,
\]
and estimator $\Omega^{({\rm min})}_{s_1,\cdots,s_d}$
,where $\{s_1,\cdots,s_d\}$ is occupation number of the state after measurement, namely, the state where
the $i$-th state is the eigenvector $|n_i \rangle $, and this  is separable measurement.
%%%%%%%%%%%%%%%%%%%% Summary and discussion %%%%%%%%%%%%%%%%%%%%%%%
\section{\label{sec:sum}summary and discussion}
In this paper, when $N$ copies of pure state which is distributed uniformly in $d$-dimensional Hilbert space are prepared, 
we showed that the separable measurement is optimal in the sense that the mean error with
respect to any measure of deviation which satisfies the condition (a) $\sim$ (c) takes the minimum.

The POVM is independent of the choice of measure $W(x)$ of deviation 
, although the estimator $\Omega^{({\rm min})}_{s_1,\cdots,s_d}$ is dependent on it.
For the case with $d=2$, $N=2$ and $\Omega=\sigma_z$, when we choose 
\begin{equation}
W(x)=\sigma^2 \sinh^2 \frac{x}{\sigma}, 
\label{exmeasdev} 
\end{equation}
 as measure of deviation,
the estimator $\omega_{s_1,s_2}$ is given by
\begin{eqnarray*}
\omega_{2,0}&=&\frac{\sigma}{4}\log\left\{
\frac{(\sigma^2-2\sigma+4)\sinh\frac{2}{\sigma}-2(\sigma-2)\cosh\frac{2}{\sigma}}
     {(\sigma^2+2\sigma+4)\sinh\frac{2}{\sigma}-2(\sigma+2)\cosh\frac{2}{\sigma}}\right\}\\
\omega_{1,1}&=&0 \\
\omega_{0,2}&=&\frac{\sigma}{4}\log\left\{
\frac{(\sigma^2+2\sigma+4)\sinh\frac{2}{\sigma}-2(\sigma+2)\cosh\frac{2}{\sigma}}
     {(\sigma^2-2\sigma+4)\sinh\frac{2}{\sigma}-2(\sigma-2)\cosh\frac{2}{\sigma}}
\right\}\\
 &=&- \omega_{2,0}   
\end{eqnarray*}
which is dependent on the parameter $\sigma$. 

The behavior of the estimator $\omega_{2,0}$ in the FIG.1 shows that the limit of $\omega_{2,0}$ as $\sigma$ approaches infinity is equal to $\frac{1}{2}$
which is obtained from the result in the previous paper
\cite{hayashi4} as is expected from the fact that the limit of measure of deviation $W(x)$
as $\sigma$ tends to infinity is equal to $x^2$.

% WinTpic
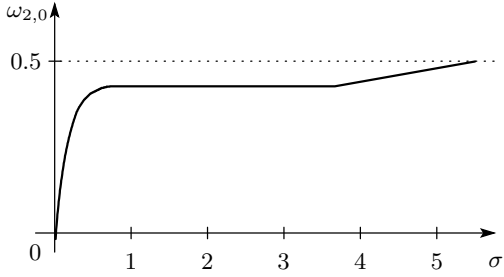
\begin{figure}[htbp]
  \begin{center}
%WinTpicVersion3.08
\unitlength 0.1in
\begin{picture}( 26.7500, 14.7000)( -0.7500,-15.0000)
% FUNC 1 0 3 0
% 9 200 200 2500 1500 300 1400 700 1400 300 400 200 200 2500 1500 0 3 0 0
% x*(ln((x*x/4-x/2+1)*(e^(2/x)-1/e^(2/x))-(x/2-1)*(e^(2/x)+1/e^(2/x)))-ln((x*x/4+x/2+1)*(e^(2/x)-1/e^(2/x))-(x/2+1)*(e^(2/x)+1/e^(2/x))))/2
\special{pn 13}%
\special{pa 306 1332}%
\special{pa 310 1282}%
\special{pa 316 1238}%
\special{pa 320 1200}%
\special{pa 326 1164}%
\special{pa 330 1130}%
\special{pa 336 1098}%
\special{pa 340 1068}%
\special{pa 346 1040}%
\special{pa 350 1014}%
\special{pa 356 990}%
\special{pa 360 966}%
\special{pa 366 942}%
\special{pa 370 922}%
\special{pa 376 900}%
\special{pa 380 880}%
\special{pa 386 862}%
\special{pa 390 844}%
\special{pa 396 826}%
\special{pa 400 810}%
\special{pa 406 794}%
\special{pa 410 778}%
\special{pa 416 764}%
\special{pa 420 750}%
\special{pa 426 736}%
\special{pa 430 724}%
\special{pa 436 712}%
\special{pa 440 700}%
\special{pa 446 688}%
\special{pa 450 678}%
\special{pa 456 668}%
\special{pa 460 658}%
\special{pa 466 648}%
\special{pa 470 638}%
\special{pa 476 630}%
\special{pa 480 622}%
\special{pa 486 614}%
\special{pa 490 606}%
\special{pa 496 600}%
\special{pa 500 592}%
\special{pa 506 586}%
\special{pa 510 580}%
\special{pa 516 574}%
\special{pa 520 568}%
\special{pa 526 562}%
\special{pa 530 556}%
\special{pa 536 552}%
\special{pa 540 546}%
\special{pa 546 542}%
\special{pa 550 538}%
\special{pa 556 534}%
\special{pa 560 530}%
\special{pa 566 526}%
\special{pa 570 522}%
\special{pa 576 518}%
\special{pa 580 514}%
\special{pa 586 512}%
\special{pa 590 508}%
\special{pa 596 506}%
\special{pa 600 502}%
\special{pa 606 500}%
\special{pa 610 496}%
\special{pa 616 494}%
\special{pa 620 492}%
\special{pa 626 490}%
\special{pa 630 486}%
\special{pa 636 484}%
\special{pa 640 482}%
\special{pa 646 480}%
\special{pa 650 478}%
\special{pa 656 476}%
\special{pa 660 474}%
\special{pa 666 472}%
\special{pa 670 470}%
\special{pa 676 470}%
\special{pa 680 468}%
\special{pa 686 466}%
\special{pa 690 464}%
\special{pa 696 464}%
\special{pa 700 462}%
\special{pa 706 460}%
\special{pa 710 458}%
\special{pa 716 458}%
\special{pa 720 456}%
\special{pa 726 456}%
\special{pa 730 454}%
\special{pa 736 452}%
\special{pa 740 452}%
\special{pa 746 450}%
\special{pa 750 450}%
\special{pa 756 448}%
\special{pa 760 448}%
\special{pa 766 446}%
\special{pa 770 446}%
\special{pa 776 444}%
\special{pa 780 444}%
\special{pa 786 444}%
\special{pa 790 442}%
\special{pa 796 442}%
\special{pa 800 440}%
\special{pa 806 440}%
\special{pa 810 440}%
\special{pa 816 438}%
\special{pa 820 438}%
\special{pa 826 438}%
\special{pa 830 436}%
\special{pa 836 436}%
\special{pa 840 436}%
\special{pa 846 434}%
\special{pa 850 434}%
\special{pa 856 434}%
\special{pa 860 434}%
\special{pa 866 432}%
\special{pa 870 432}%
\special{pa 876 432}%
\special{pa 880 430}%
\special{pa 886 430}%
\special{pa 890 430}%
\special{pa 896 430}%
\special{pa 900 430}%
\special{pa 906 428}%
\special{pa 910 428}%
\special{pa 916 428}%
\special{pa 920 428}%
\special{pa 926 426}%
\special{pa 930 426}%
\special{pa 936 426}%
\special{pa 940 426}%
\special{pa 946 426}%
\special{pa 950 424}%
\special{pa 956 424}%
\special{pa 960 424}%
\special{pa 966 424}%
\special{pa 970 424}%
\special{pa 976 424}%
\special{pa 980 422}%
\special{pa 986 422}%
\special{pa 990 422}%
\special{pa 996 422}%
\special{pa 1000 422}%
\special{pa 1006 422}%
\special{pa 1010 422}%
\special{pa 1016 420}%
\special{pa 1020 420}%
\special{pa 1026 420}%
\special{pa 1030 420}%
\special{pa 1036 420}%
\special{pa 1040 420}%
\special{pa 1046 420}%
\special{pa 1050 420}%
\special{pa 1056 418}%
\special{pa 1060 418}%
\special{pa 1066 418}%
\special{pa 1070 418}%
\special{pa 1076 418}%
\special{pa 1080 418}%
\special{pa 1086 418}%
\special{pa 1090 418}%
\special{pa 1096 418}%
\special{pa 1100 416}%
\special{pa 1106 416}%
\special{pa 1110 416}%
\special{pa 1116 416}%
\special{pa 1120 416}%
\special{pa 1126 416}%
\special{pa 1130 416}%
\special{pa 1136 416}%
\special{pa 1140 416}%
\special{pa 1146 416}%
\special{pa 1150 414}%
\special{pa 1156 414}%
\special{pa 1160 414}%
\special{pa 1166 414}%
\special{pa 1170 414}%
\special{pa 1176 414}%
\special{pa 1180 414}%
\special{pa 1186 414}%
\special{pa 1190 414}%
\special{pa 1196 414}%
\special{pa 1200 414}%
\special{pa 1206 414}%
\special{pa 1210 414}%
\special{pa 1216 414}%
\special{pa 1220 412}%
\special{pa 1226 412}%
\special{pa 1230 412}%
\special{pa 1236 412}%
\special{pa 1240 412}%
\special{pa 1246 412}%
\special{pa 1250 412}%
\special{pa 1256 412}%
\special{pa 1260 412}%
\special{pa 1266 412}%
\special{pa 1270 412}%
\special{pa 1276 412}%
\special{pa 1280 412}%
\special{pa 1286 412}%
\special{pa 1290 412}%
\special{pa 1296 412}%
\special{pa 1300 412}%
\special{pa 1306 410}%
\special{pa 1310 410}%
\special{pa 1316 410}%
\special{pa 1320 410}%
\special{pa 1326 410}%
\special{pa 1330 410}%
\special{pa 1336 410}%
\special{pa 1340 410}%
\special{pa 1346 410}%
\special{pa 1350 410}%
\special{pa 1356 410}%
\special{pa 1360 410}%
\special{pa 1366 410}%
\special{pa 1370 410}%
\special{pa 1376 410}%
\special{pa 1380 410}%
\special{pa 1386 410}%
\special{pa 1390 410}%
\special{pa 1396 410}%
\special{pa 1400 410}%
\special{pa 1406 410}%
\special{pa 1410 410}%
\special{pa 1416 408}%
\special{pa 1420 408}%
\special{pa 1426 408}%
\special{pa 1430 408}%
\special{pa 1436 408}%
\special{pa 1440 408}%
\special{pa 1446 408}%
\special{pa 1450 408}%
\special{pa 1456 408}%
\special{pa 1460 408}%
\special{pa 1466 408}%
\special{pa 1470 408}%
\special{pa 1476 408}%
\special{pa 1480 408}%
\special{pa 1486 408}%
\special{pa 1490 408}%
\special{pa 1496 408}%
\special{pa 1500 408}%
\special{pa 1506 408}%
\special{pa 1510 408}%
\special{pa 1516 408}%
\special{pa 1520 408}%
\special{pa 1526 408}%
\special{pa 1530 408}%
\special{pa 1536 408}%
\special{pa 1540 408}%
\special{pa 1546 408}%
\special{pa 1550 408}%
\special{pa 1556 408}%
\special{pa 1560 408}%
\special{pa 1566 408}%
\special{pa 1570 408}%
\special{pa 1576 408}%
\special{pa 1580 406}%
\special{pa 1586 406}%
\special{pa 1590 406}%
\special{pa 1596 406}%
\special{pa 1600 406}%
\special{pa 1606 406}%
\special{pa 1610 406}%
\special{pa 1616 406}%
\special{pa 1620 406}%
\special{pa 1626 406}%
\special{pa 1630 406}%
\special{pa 1636 406}%
\special{pa 1640 406}%
\special{pa 1646 406}%
\special{pa 1650 406}%
\special{pa 1656 406}%
\special{pa 1660 406}%
\special{pa 1666 406}%
\special{pa 1670 406}%
\special{pa 1676 406}%
\special{pa 1680 406}%
\special{pa 1686 406}%
\special{pa 1690 406}%
\special{pa 1696 406}%
\special{pa 1700 406}%
\special{pa 1706 406}%
\special{pa 1710 406}%
\special{pa 1716 406}%
\special{pa 1720 406}%
\special{pa 1726 406}%
\special{pa 1730 406}%
\special{pa 1736 406}%
\special{pa 1740 406}%
\special{pa 1746 406}%
\special{pa 1750 406}%
\special{pa 1756 406}%
\special{pa 1760 406}%
\special{pa 1766 406}%
\special{pa 1770 406}%
\special{pa 1776 406}%
\special{pa 1780 406}%
\special{pa 1786 406}%
\special{pa 1790 406}%
\special{pa 1796 406}%
\special{pa 1800 406}%
\special{pa 1806 406}%
\special{pa 1810 406}%
\special{pa 1816 406}%
\special{pa 1820 406}%
\special{pa 1826 406}%
\special{pa 1830 406}%
\special{pa 1836 406}%
\special{pa 1840 404}%
\special{pa 1846 404}%
\special{pa 1850 404}%
\special{pa 1856 404}%
\special{pa 1860 404}%
\special{pa 1866 404}%
\special{pa 1870 404}%
\special{pa 1876 404}%
\special{pa 1880 404}%
\special{pa 1886 404}%
\special{pa 1890 404}%
\special{pa 1896 404}%
\special{pa 1900 404}%
\special{pa 1906 404}%
\special{pa 1910 404}%
\special{pa 1916 404}%
\special{pa 1920 404}%
\special{pa 1926 404}%
\special{pa 1930 404}%
\special{pa 1936 404}%
\special{pa 1940 404}%
\special{pa 1946 404}%
\special{pa 1950 404}%
\special{pa 1956 404}%
\special{pa 1960 404}%
\special{pa 1966 404}%
\special{pa 1970 404}%
\special{pa 1976 404}%
\special{pa 1980 404}%
\special{pa 1986 404}%
\special{pa 1990 404}%
\special{pa 1996 404}%
\special{pa 2000 404}%
\special{pa 2006 404}%
\special{pa 2010 404}%
\special{pa 2016 404}%
\special{pa 2020 404}%
\special{pa 2026 404}%
\special{pa 2030 404}%
\special{pa 2036 404}%
\special{pa 2040 404}%
\special{pa 2046 404}%
\special{pa 2050 404}%
\special{pa 2056 404}%
\special{pa 2060 404}%
\special{pa 2066 404}%
\special{pa 2070 404}%
\special{pa 2076 404}%
\special{pa 2080 404}%
\special{pa 2086 404}%
\special{pa 2090 404}%
\special{pa 2096 404}%
\special{pa 2100 404}%
\special{pa 2106 404}%
\special{pa 2110 404}%
\special{pa 2116 404}%
\special{pa 2120 404}%
\special{pa 2126 404}%
\special{pa 2130 404}%
\special{pa 2136 404}%
\special{pa 2140 404}%
\special{pa 2146 404}%
\special{pa 2150 404}%
\special{pa 2156 404}%
\special{pa 2160 404}%
\special{pa 2166 404}%
\special{pa 2170 404}%
\special{pa 2176 404}%
\special{pa 2180 404}%
\special{pa 2186 404}%
\special{pa 2190 404}%
\special{pa 2196 404}%
\special{pa 2200 404}%
\special{pa 2206 404}%
\special{pa 2210 404}%
\special{pa 2216 404}%
\special{pa 2220 404}%
\special{pa 2226 404}%
\special{pa 2230 404}%
\special{pa 2236 404}%
\special{pa 2240 404}%
\special{pa 2246 404}%
\special{pa 2250 404}%
\special{pa 2256 404}%
\special{pa 2260 404}%
\special{pa 2266 404}%
\special{pa 2270 404}%
\special{pa 2276 404}%
\special{pa 2280 404}%
\special{pa 2286 404}%
\special{pa 2290 404}%
\special{pa 2296 404}%
\special{pa 2300 404}%
\special{pa 2306 404}%
\special{pa 2310 404}%
\special{pa 2316 404}%
\special{pa 2320 404}%
\special{pa 2326 404}%
\special{pa 2330 404}%
\special{pa 2336 404}%
\special{pa 2340 404}%
\special{pa 2346 404}%
\special{pa 2350 404}%
\special{pa 2356 404}%
\special{pa 2360 404}%
\special{pa 2366 402}%
\special{pa 2370 402}%
\special{pa 2376 402}%
\special{pa 2380 402}%
\special{pa 2386 402}%
\special{pa 2390 402}%
\special{pa 2396 402}%
\special{pa 2400 402}%
\special{pa 2406 402}%
\special{pa 2410 402}%
\special{pa 2416 402}%
\special{pa 2420 402}%
\special{pa 2426 402}%
\special{pa 2430 402}%
\special{pa 2436 402}%
\special{pa 2440 402}%
\special{pa 2446 402}%
\special{pa 2450 402}%
\special{pa 2456 402}%
\special{pa 2460 402}%
\special{pa 2466 402}%
\special{pa 2470 402}%
\special{pa 2476 402}%
\special{pa 2480 402}%
\special{pa 2486 402}%
\special{pa 2490 402}%
\special{pa 2496 402}%
\special{pa 2500 402}%
\special{sp}%
% VECTOR 2 0 3 0
% 4 300 1400 300 100 200 1300 2600 1300
% 
\special{pn 8}%
\special{pa 300 1400}%
\special{pa 300 100}%
\special{fp}%
\special{sh 1}%
\special{pa 300 100}%
\special{pa 280 168}%
\special{pa 300 154}%
\special{pa 320 168}%
\special{pa 300 100}%
\special{fp}%
\special{pa 200 1300}%
\special{pa 2600 1300}%
\special{fp}%
\special{sh 1}%
\special{pa 2600 1300}%
\special{pa 2534 1280}%
\special{pa 2548 1300}%
\special{pa 2534 1320}%
\special{pa 2600 1300}%
\special{fp}%
% LINE 2 0 3 0
% 2 700 1270 700 1330
% 
\special{pn 8}%
\special{pa 700 1270}%
\special{pa 700 1330}%
\special{fp}%
% LINE 2 0 3 0
% 2 1100 1270 1100 1330
% 
\special{pn 8}%
\special{pa 1100 1270}%
\special{pa 1100 1330}%
\special{fp}%
% LINE 2 0 3 0
% 2 1500 1270 1500 1330
% 
\special{pn 8}%
\special{pa 1500 1270}%
\special{pa 1500 1330}%
\special{fp}%
% LINE 2 0 3 0
% 2 1900 1270 1900 1330
% 
\special{pn 8}%
\special{pa 1900 1270}%
\special{pa 1900 1330}%
\special{fp}%
% LINE 2 0 3 0
% 2 2300 1270 2300 1330
% 
\special{pn 8}%
\special{pa 2300 1270}%
\special{pa 2300 1330}%
\special{fp}%
% LINE 2 0 3 0
% 2 270 400 330 400
% 
\special{pn 8}%
\special{pa 270 400}%
\special{pa 330 400}%
\special{fp}%
% LINE 2 2 3 0
% 2 330 400 2600 400
% 
\special{pn 8}%
\special{pa 330 400}%
\special{pa 2600 400}%
\special{dt 0.045}%
% STR 2 0 3 0
% 3 200 1300 200 1400 5 0
% $0$
\put(2.0000,-14.0000){\makebox(0,0){$0$}}%
% STR 2 0 3 0
% 3 2610 1350 2610 1450 5 0
% $\sigma$
\put(26.1000,-14.5000){\makebox(0,0){$\sigma$}}%
% STR 2 0 3 0
% 3 50 100 50 200 2 0
% $\omega_{2,0}$
\put(0.5000,-2.0000){\makebox(0,0)[lb]{$\omega_{2,0}$}}%
% STR 2 0 3 0
% 3 150 300 150 400 5 0
% $0.5$
\put(1.5000,-4.0000){\makebox(0,0){$0.5$}}%
% STR 2 0 3 0
% 3 700 1350 700 1450 5 0
% $1$
\put(7.0000,-14.5000){\makebox(0,0){$1$}}%
% STR 2 0 3 0
% 3 1100 1350 1100 1450 5 0
% $2$
\put(11.0000,-14.5000){\makebox(0,0){$2$}}%
% STR 2 0 3 0
% 3 1500 1350 1500 1450 5 0
% $3$
\put(15.0000,-14.5000){\makebox(0,0){$3$}}%
% STR 2 0 3 0
% 3 1900 1350 1900 1450 5 0
% $4$
\put(19.0000,-14.5000){\makebox(0,0){$4$}}%
% STR 2 0 3 0
% 3 2300 1350 2300 1450 5 0
% $5$
\put(23.0000,-14.5000){\makebox(0,0){$5$}}%
\end{picture}%

  \end{center}
  \caption{The estimator $\omega_{2,0}$ as a function of parameter $\sigma$ in measure(\ref{exmeasdev}) of deviation. When $\sigma$ goes to infinity, the estimator $\omega_{2,0}$ approaches $\frac{1}{2}$ which is equal to the estimator for the case with $d=2$, $N=2$ and the observable $\sigma_z$ in the previous paper \cite{hayashi4}.}%
\end{figure}

We assumed that the measure of deviation satisfies the conditions (a) $\sim$ (c). These conditions are sufficient conditions for existence of unique value $\Omega^{({\rm min})}_{s_1,\cdots,s_d}$ where the function $w_{s_1,\cdots.s_d}(\omega_a)$ defined by the equation (\ref{deffuw}) takes the minimum.
Even if conditions (b) and (c) are replaced with weaker condition (b')
\begin{itemize}
\item[(b')] the function $W(x)$ is continuous and monotonically decreasing (increasing) function if $x$ is negative (positive),  
\end{itemize}
existence of minimum value of the function $w_{s_1,\cdots.s_d}(\omega_a)$ is shown as
follows. 
Because the function $w_{s_1,\cdots.s_d}(\omega_a)$ is continuous, in the finite region between the smallest eigenvalue $\lambda_{\rm min}$  and the largest eigenvalue $\lambda_{\rm max}$
of observable $\Omega$
, there is, at least, one value 
${\Omega'}^{({\rm min})}_a$ at which
the function $w_{s_1,\cdots.s_d}(\omega_a)$ takes the minimum.
For the $\omega_a$ which is smaller than the smallest eigenvalue $\lambda_{\rm min}$ of observable $\Omega$, because
$\omega_a-\sum_{n=1}^d\lambda_{n}\xi_n^2$ is negative;
\[
\omega_a-\sum_{n=1}^d\lambda_{n}\xi_n^2 \le \omega_a - \lambda_{\rm min} <0,
\]
the measure of deviation $W(\omega_a-\sum_{n=1}^d\lambda_{n}\xi_n^2)$ is monotonically decreasing function of variable $\omega_a$ and the function $w_{s_1,\cdots.s_d}(\omega_a)$ is monotonically decreasing function of variable $\omega_a$, too.
Similarly, for the $\omega_a$ which is larger than the largest eigenvalue $\lambda_{\rm max}$, $w_{s_1,\cdots.s_d}(\omega_a)$ is monotonically increasing function of variable $\omega_a$.
 Hence, at $\omega_a={\Omega'}^{({\rm min})}_a$
this function becomes minimum over the interval $(-\infty, +\infty)$. 

In the case with the condition (b') in stead of the conditions (b) and (c)  , we  may have some points at which the function $w_{s_1,\cdots.s_d}(\omega_a)$
take the minimum and cannot uniquely determine
estimator corresponding to each element of projection operator.

We investigated the case where pure state is distributed uniformly. Because of this
assumption, integrand in the equation (\ref{exdefdev}) except for the factor
$\prod_{i,j}^Nc_{n_i}c^{\ast}_{m_i}$ is independent of complex argument $\varphi_i$ of coefficient
$c_n$ and we obtained the equation (\ref{exdefdev2}).  More generally, for the case where
pure states are distributed by the probability $p(|c_1|,\cdots,|c_N|)$, the mean error
$\langle \Delta' \rangle$ becomes
\begin{eqnarray*}
\langle \Delta' \rangle&=&\sum_a\int\prod_{n=1}^ddc_ndc_n^{\ast}
\delta\left(\sum_{n=1}^d|c_n|^2-1\right)\\
&\times&p(|c_1|,\cdots,|c_N|)\sum_{n_1=1}^d\cdots\sum_{n_N=1}^d 
        \sum_{m_1=1}^d\cdots\sum_{m_N=1}^d\\
&\times& \prod_{i,j}^Nc_{n_i}c^{\ast}_{m_i} 
{\rm Tr}[|n_1\rangle\langle m_1| \otimes \cdots  \otimes |n_N\rangle\langle m_N| E_a]\\
&\times&W(\omega_a-\sum_{n=1}^d\lambda_{n}|c_n|^2),
\end{eqnarray*}
and we obtain equation (\ref{exdefdev2}) with $w'_{s_1,\cdots,s_d}(\omega_a)$
\begin{eqnarray*}
w'_{s_1,\cdots,s_d}(\omega_a)&=&(2\pi)^d\int_0^{\infty}\prod_{n=1}^dd\xi_n
\delta\left(\sum_{n=1}^d\xi_n^2-1\right)\\
&\times&p(\xi_1,\cdots,\xi_N)\prod_{n=1}^d\xi_n^{2s_n+1}\\
&\times&W\left(\omega_a-\sum_{n=1}^d\lambda_{n}\xi_n^2\right),
\end{eqnarray*}
instead of $w_{s_1,\cdots,s_d}(\omega_a)$.
In the same manner as in the previous section, it is shown that the separable measurement on each copy is optimal.
For example, we consider measurement of observable $\sigma_z$ on 
$2$-dimensional Hilbert space. 
Using  polar coordinate $(\theta, \varphi)$ on Bloch sphere,
any pure state $|\phi\rangle$ is described in the form,
\[
|\phi\rangle=\cos\frac{\theta}{2}|0\rangle + 
 e^{i\varphi}\sin\frac{\theta}{2}|1\rangle,
\]
up to phase factor, where $|0\rangle$ and $|1\rangle$ are eigenvectors of $\sigma_z$
\[
\sigma_z|0 \rangle= +1|0 \rangle,\;\;\sigma_z|1 \rangle= -1|1 \rangle.
\]
If the probability density for distribution of pure state is independent
of azimuthal angle $\varphi$, namely, if the distribution of pure state is symmetric under the rotation around $z$ axis, separable measurement on each copy makes the mean error minimum and is one of the optimal measurements.

%\begin{acknowledgments}
%We wish to acknowledge the support of the author community in using
%REV\TeX{}, offering suggestions and encouragement, testing new versions,
%\end{acknowledgments}

\newpage %Just because of unusual number of tables stacked at end
\bibliography{msrmntf}% Produces the bibliography via BibTeX.

\begin{thebibliography}{18}
\expandafter\ifx\csname natexlab\endcsname\relax\def\natexlab#1{#1}\fi
\expandafter\ifx\csname bibnamefont\endcsname\relax
  \def\bibnamefont#1{#1}\fi
\expandafter\ifx\csname bibfnamefont\endcsname\relax
  \def\bibfnamefont#1{#1}\fi
\expandafter\ifx\csname citenamefont\endcsname\relax
  \def\citenamefont#1{#1}\fi
\expandafter\ifx\csname url\endcsname\relax
  \def\url#1{\texttt{#1}}\fi
\expandafter\ifx\csname urlprefix\endcsname\relax\def\urlprefix{URL }\fi
\providecommand{\bibinfo}[2]{#2}
\providecommand{\eprint}[2][]{\url{#2}}

\bibitem[{\citenamefont{Helstrom}(1976)}]{Helst}
\bibinfo{author}{\bibfnamefont{C.~W.} \bibnamefont{Helstrom}},
  \emph{\bibinfo{title}{Quantum Detection and Estimation Theory}}
  (\bibinfo{publisher}{Academic Press, New York}, \bibinfo{year}{1976}).

\bibitem[{\citenamefont{Holevo}(1982)}]{Holevo}
\bibinfo{author}{\bibfnamefont{A.~S.} \bibnamefont{Holevo}},
  \emph{\bibinfo{title}{Probabilistic and Statistical Aspect of Quantum
  Theory}} (\bibinfo{publisher}{North-Holland, Amsterdam},
  \bibinfo{year}{1982}).

\bibitem[{\citenamefont{{S.Massar and S.Popescu}}(1995)}]{Massar95}
\bibinfo{author}{\bibnamefont{{S.Massar and S.Popescu}}},
  \bibinfo{journal}{Phys.\ Rev.\ Lett.} \textbf{\bibinfo{volume}{74}},
  \bibinfo{pages}{1259} (\bibinfo{year}{1995}).

\bibitem[{\citenamefont{{R.Derka, V.Bu$\breve{\rm z}$ek, and
  A.K.Ekert}}(1998)}]{Derka98}
\bibinfo{author}{\bibnamefont{{R.Derka, V.Bu$\breve{\rm z}$ek, and
  A.K.Ekert}}}, \bibinfo{journal}{Phys.\ Rev.\ Lett.}
  \textbf{\bibinfo{volume}{80}}, \bibinfo{pages}{1571} (\bibinfo{year}{1998}).

\bibitem[{\citenamefont{{Bru{\ss} and Chiara Macchiavello}}(1999)}]{Bruss99}
\bibinfo{author}{\bibnamefont{{Bru{\ss} and Chiara Macchiavello}}},
  \bibinfo{journal}{Phys.\ Lett.\ A} \textbf{\bibinfo{volume}{253}},
  \bibinfo{pages}{249} (\bibinfo{year}{1999}).

\bibitem[{\citenamefont{{Chiribella, G. M. D'Ariano, P. Perinotti, and M. F.
  Sacchi}}(2004)}]{chirib042}
\bibinfo{author}{\bibnamefont{{Chiribella, G. M. D'Ariano, P. Perinotti, and M.
  F. Sacchi}}}, \bibinfo{journal}{Phys.\ Rev.} \textbf{\bibinfo{volume}{A 70}},
  \bibinfo{pages}{062105} (\bibinfo{year}{2004}).

\bibitem[{\citenamefont{{A. Hayashi, M. Horibe and T.
  Hashimoto}}(2005{\natexlab{a}})}]{hayashi1}
\bibinfo{author}{\bibnamefont{{A. Hayashi, M. Horibe and T. Hashimoto}}},
  \bibinfo{journal}{Phys.\ Rev.} \textbf{\bibinfo{volume}{A 72}},
  \bibinfo{pages}{032325} (\bibinfo{year}{2005}{\natexlab{a}}).

\bibitem[{\citenamefont{{E.Bagan, M.Baig, and
  R.Mu\~{n}oz-Tapia}}(2001{\natexlab{a}})}]{Bag101}
\bibinfo{author}{\bibnamefont{{E.Bagan, M.Baig, and R.Mu\~{n}oz-Tapia}}},
  \bibinfo{journal}{Phys.\ Rev.} \textbf{\bibinfo{volume}{A 64}},
  \bibinfo{pages}{022305} (\bibinfo{year}{2001}{\natexlab{a}}).

\bibitem[{\citenamefont{{E.Bagan, M.Baig, and
  R.Mu\~{n}oz-Tapia}}(2001{\natexlab{b}})}]{Bag201}
\bibinfo{author}{\bibnamefont{{E.Bagan, M.Baig, and R.Mu\~{n}oz-Tapia}}},
  \bibinfo{journal}{Phys.\ Rev.\ Lett.} \textbf{\bibinfo{volume}{87}},
  \bibinfo{pages}{257903} (\bibinfo{year}{2001}{\natexlab{b}}).

\bibitem[{\citenamefont{{G. Chiribella, G. M. D'Ariano, P. Perinotti, and M. F.
  Sacchi}}(2004)}]{chirib041}
\bibinfo{author}{\bibnamefont{{G. Chiribella, G. M. D'Ariano, P. Perinotti, and
  M. F. Sacchi}}}, \bibinfo{journal}{Phys.\ Rev.\ Lett.}
  \textbf{\bibinfo{volume}{93}}, \bibinfo{pages}{180503}
  (\bibinfo{year}{2004}).

\bibitem[{\citenamefont{{A. Hayashi, M. Horibe and T.
  Hashimoto}}(2005{\natexlab{b}})}]{hayashi2}
\bibinfo{author}{\bibnamefont{{A. Hayashi, M. Horibe and T. Hashimoto}}},
  \bibinfo{journal}{Phys.\ Rev.} \textbf{\bibinfo{volume}{A 72}},
  \bibinfo{pages}{052306} (\bibinfo{year}{2005}{\natexlab{b}}).

\bibitem[{\citenamefont{{A. Hayashi, M. Horibe and T.
  Hashimoto}}(2006{\natexlab{a}})}]{hayashi3}
\bibinfo{author}{\bibnamefont{{A. Hayashi, M. Horibe and T. Hashimoto}}},
  \bibinfo{journal}{Phys.\ Rev.} \textbf{\bibinfo{volume}{A 73}},
  \bibinfo{pages}{012328} (\bibinfo{year}{2006}{\natexlab{a}}).

\bibitem[{\citenamefont{Hayashi}(1998)}]{Mhayashi}
\bibinfo{author}{\bibfnamefont{M.}~\bibnamefont{Hayashi}},
  \bibinfo{journal}{J.\ Phys.\ A} \textbf{\bibinfo{volume}{31}},
  \bibinfo{pages}{4633} (\bibinfo{year}{1998}).

\bibitem[{\citenamefont{{E.Bagan, M.A.Ballester, R.Mu{\~n}oz-Tapia and
  O.Romero-Isart}}(2005)}]{Bagan05}
\bibinfo{author}{\bibnamefont{{E.Bagan, M.A.Ballester, R.Mu{\~n}oz-Tapia and
  O.Romero-Isart}}}, \bibinfo{journal}{Phys.\ Rev.\ Lett}
  \textbf{\bibinfo{volume}{95}}, \bibinfo{pages}{110504}
  (\bibinfo{year}{2005}).

\bibitem[{\citenamefont{{Manuel A. Ballester}}(2004)}]{Ball04}
\bibinfo{author}{\bibnamefont{{Manuel A. Ballester}}}, \bibinfo{journal}{Phys.\
  Rev.} \textbf{\bibinfo{volume}{A 70}}, \bibinfo{pages}{032310}
  (\bibinfo{year}{2004}).

\bibitem[{\citenamefont{{V.Giovannetti, S.Lloyd and l.macone}}(2006)}]{Giov06}
\bibinfo{author}{\bibnamefont{{V.Giovannetti, S.Lloyd and l.macone}}},
  \bibinfo{journal}{Phys.\ Rev.\ Lett} \textbf{\bibinfo{volume}{96}},
  \bibinfo{pages}{010401} (\bibinfo{year}{2006}).

\bibitem[{\citenamefont{{G. M. D'Ariano, V. Giovannetti and P.
  Perinotri}}(2006)}]{DAGiPer}
\bibinfo{author}{\bibnamefont{{G. M. D'Ariano, V. Giovannetti and P.
  Perinotri}}}, \bibinfo{journal}{J.\ Math.\ Phys.}
  \textbf{\bibinfo{volume}{47}}, \bibinfo{pages}{022102}
  (\bibinfo{year}{2006}).

\bibitem[{\citenamefont{{A. Hayashi, M. Horibe and T.
  Hashimoto}}(2006{\natexlab{b}})}]{hayashi4}
\bibinfo{author}{\bibnamefont{{A. Hayashi, M. Horibe and T. Hashimoto}}},
  \bibinfo{journal}{Phys.\ Rev.} \textbf{\bibinfo{volume}{A 73}},
  \bibinfo{pages}{062322} (\bibinfo{year}{2006}{\natexlab{b}}).

\end{thebibliography}

\end{document}